\documentclass[12pt,twoside,a4paper]{article}
\usepackage{graphicx}
\usepackage{amsmath}
\usepackage{color}

\begin{document}
\thispagestyle{empty} 
\title{
\vskip-3cm
{\baselineskip14pt
\centerline{\normalsize DESY 20-179 \hfill ISSN 0418--9833}
\centerline{\normalsize MITP/20--074 \hfill} 
\centerline{\normalsize February 2021 \hfill}} 
\vskip1.5cm
\boldmath
{\bf 
Cross sections of inclusive $\psi(2S)$ and 
\\ 
$X(3872)$ production from 
\\
$b$-hadron decays in $pp$ collisions and 
\\ 
comparison with ATLAS, CMS, and LHCb data}
\unboldmath
\author{
%M. Benzke$^1$,
B.~A.~Kniehl$^1$, 
G.~Kramer$^1$, 
I.~Schienbein$^2$ 
and H.~Spiesberger$^3$
\vspace{2mm} \\
\normalsize{
  $^1$ II. Institut f\"ur Theoretische
  Physik, Universit\"at Hamburg,
}\\ 
\normalsize{
  Luruper Chaussee 149, D-22761 Hamburg, Germany
} \vspace{2mm}\\
\normalsize{
  $^2$ Laboratoire de Physique Subatomique et de Cosmologie,
} \\ 
\normalsize{
  Universit\'e Joseph Fourier Grenoble 1,
}\\
\normalsize{
  CNRS/IN2P3, Institut National Polytechnique de Grenoble,
}\\
\normalsize{
  53 avenue des Martyrs, F-38026 Grenoble, France
} \vspace{2mm}\\
\normalsize{
  $^3$ PRISMA$^+$ Cluster of Excellence, Institut f\"ur Physik,
  Johannes-Gutenberg-Universit\"at,
}\\ 
\normalsize{
  Staudinger Weg 7, D-55099 Mainz, Germany,}\\
\vspace{2mm} \\}}
%\date{01.04.2019}
\maketitle

%%%%%%%%%%%%%%%%%%%%%%%%%%%%%%%%%%%%%%%%%%%%%%%%%%%%%%%%%%%%%%%%%%%%%
\begin{abstract}
\medskip
\noindent 
We study the cross sections for the inclusive production of 
$\psi(2S)$ and $X(3872)$ hadrons in $pp$ collisions at the LHC 
at two different center-of-mass energies and compare them
with experimental data obtained by the ATLAS, CMS, and LHCb 
Collaborations.   
\end{abstract}
\clearpage
%%%%%%%%%%%%%%%%%%%%%%%%%%%%%%%%%%%%%%%%%%%%%%%%%%%%%%%%%%%%%%%%%%%%%
%%%%%%%%%%%%%%%%%%%%%%%%%%%%%%%%%%%%%%%%%%%%%%%%%%%%%%%%%%%%%%%%%%%%%

\section{Intoduction}

Some time ago, Paolo Bolzoni and two of us calculated the cross 
section for the inclusive production of $J/\psi$ and $\psi(2S)$ 
mesons originating from decays of $B$ mesons produced in $p\bar{p}$ 
collisions with center-of-mass energy $\sqrt{S} = 1.96$~TeV at the 
Fermilab Tevatron and in $pp$ collisions with $\sqrt{S} = 7$~TeV 
at the CERN LHC at next-to-leading order (NLO) in the framework of 
the general-mass variable-flavor number scheme (GM-VFNS) in connection
with nonrelativistic-QCD (NRQCD) factorization \cite{Bolzoni:2013tca}.
In Ref.~\cite{Bolzoni:2013tca}, the transverse momentum 
($p_T$) distributions of such nonprompt $J/\psi$ mesons measured 
by the 
CDF~II \cite{Acosta:2004yw,Aaltonen:2009dm}, 
CMS \cite{Khachatryan:2010yr,Chatrchyan:2011kc}, 
LHCb \cite{Aaij:2011jh,Aaij:2012ag}, 
ATLAS \cite{Aad:2011sp}, and 
ALICE \cite{Abelev:2012gx}
Collaborations were found to be very well described by our 
predictions, with respect to both absolute normalization and line 
shape. Similarly, the $p_T$ distributions of $\psi(2S)$ nonprompt 
production measured by 
CDF~II \cite{Aaltonen:2009dm}, 
CMS \cite{Chatrchyan:2011kc}, and 
LHCb \cite{Aaij:2012ag} were rather well described by our 
calculations. 

In 2003, a narrow charmonium-like state was discovered in exclusive
$B^+ \to J/\psi K^+\pi^+\pi^-$ decays by the BELLE Collaboration 
\cite{Choi:2003ue}. The subsequent development was described in 
detail in publications by CMS \cite{Chatrchyan:2013cld} and ATLAS 
\cite{Aaboud:2016vzw}, where the first measurements of $X(3872)$ 
at the LHC were reported, and in two review articles 
\cite{Ali:2017jda,Brambilla:2019esw}.

In 2017, ATLAS published measurements of the $\psi(2S)$ and $X(3872)$ 
cross sections in $pp$ collisions at $\sqrt{S} = 8$~TeV, for both 
prompt and nonprompt production \cite{Aaboud:2016vzw}. Both the 
$\psi(2S)$ and $X(3872)$ hadrons were detected via their decays to
$J/\psi\pi^+\pi^-$. The experimental results for prompt 
production were found to be in agreement with predictions of 
nonrelativistic QCD (NRQCD) factorization
\cite{Butenschoen:2013pxa,Meng:2013gga,Butenschoen:2019npa}. 
The $\psi(2S)$ nonprompt production cross section was compared 
with FONLL predictions \cite{Cacciari:2012ny} and also found to 
agree very well with the ATLAS data \cite{Aaboud:2016vzw}. 

Also the cross section of $X(3872)$ nonprompt production measured 
by ATLAS \cite{Aaboud:2016vzw} was compared with theory.
Specifically, the nonprompt $\psi(2S)$ production cross section 
evaluated in the FONLL scheme was rescaled by the ratio of branching
fractions\footnote{%
  Notice that Eq.~\eqref{eq:rb} implies a summation over the various $B$-hadron
  species and that the tacitly assumed universality of $R_B$ is based on the
  assumption that the $B$-hadron fragmentation functions (FFs) are process
  independent, as they should by the factorization theorem
  \cite{Collins:1998rz}.} 
\begin{equation}
R_B= 
\frac{{\rm Br}(B\to X(3872) + X) 
      {\rm Br}(X(3872) \to J/\psi \pi^+\pi^-)}
     {{\rm Br}(B \to \psi(2S) +X) 
      {\rm Br}(\psi(2S) \to J/\psi \pi^+\pi^-)}\,,
\label{eq:rb}
\end{equation} 
which was evaluated using the result
${\rm Br}(B \to X(3872) + X) {\rm Br}(X(3872) \to J/\psi \pi^+\pi^-) 
= (1.9 \pm 0.8) \times 10^{-4}$ extracted in 
Ref.~\cite{Artoisenet:2009wk} from Tevatron data \cite{Bauer:2004bc} 
and the values ${\rm Br}(B \to \psi(2S) + X) = (3.07 \pm 0.21) \times 
10^{-3}$ and ${\rm Br}(\psi(2S) \to J/\psi\pi^+\pi^-) = 0.3446 \pm 
0.0030$ quoted by the Particle Data Group \cite{Zyla:2020zbs} to give
\begin{equation}
R_B^{\text{\cite{Artoisenet:2009wk}}} = 0.18 \pm 0.08 \,. 
\label{rbhi}
\end{equation} 
This led to an overestimation of the ATLAS data by a large factor, 
increasing with $p_T$ from about 4 to about 8 \cite{Aaboud:2016vzw}.

We observe that the value of $R_B$ in Eq.~(\ref{rbhi}) is a factor 
of 5 larger than the one measured by ATLAS in a two-lifetime fit 
\cite{Aaboud:2016vzw},
\begin{equation}
R_B^{\text{2L}} = (3.57 \pm 0.348) \times 10^{-2}.
\label{eq:rb2l}
\end{equation}
If this lower value had been used instead of the larger value 
based on Ref.~\cite{Artoisenet:2009wk}, the differential cross 
section $d\sigma/dp_T$ of $X(3872)$ nonprompt production 
calculated in the FONLL framework would have been in approximate
agreement with the ATLAS measurement \cite{Aaboud:2016vzw}. We 
are not aware of an explanation for the discrepancy between the 
two values of $R_B$. 

The $\psi(2S)$ and $X(3872)$ processes considered here differ 
in the mechanisms of both production from $B$-hadron decay and
decay to $J/\psi\pi^+\pi^-$. The $\psi(2S)$ 
meson is a pure $c\bar{c}$ state, whereas the $X(3872)$ hadron is 
believed to be a quantum-mechanical mixture of the pure $c\bar{c}$ 
state $\chi_{1c}(2P)$ and a $D^0 \bar{D}^{*0}$ molecule, which 
is predominantly produced via its $\chi_{1c}(2P)$ component 
\cite{Butenschoen:2013pxa,Meng:2013gga,Butenschoen:2019npa}.
These differences are reflected in the fact that $R_B$ is so different
from unity.
The value of $R_B$ could be predicted from purely theoretical considerations,
{\it e.g.}, on the basis of effective field theories derived from QCD in
combination with hadron models.
However, this would reach beyond the scope of this work.
Instead, we adopt here a more heuristic approach to $R_B$, based on the
interpretation of experimental data.

In the present work, we shall present the results of an alternative 
calculation of $d\sigma/dp_T$ for $\psi(2S)$ nonprompt production 
in the kinematic range relevant for the CMS \cite{Chatrchyan:2013cld} and ATLAS
\cite{Aaboud:2016vzw} measurements, based on the GM-VFNS. The calculation of 
$d\sigma/dp_T$ for a fixed rapidity ($y$) range was described in 
detail in Ref.~\cite{Bolzoni:2013tca}, which is based on the earlier 
analysis \cite{Kniehl:1999vf} based on the zero-mass 
variable-flavor-number scheme (ZM-VFNS), where bottom is included
among the massless quark flavors. An alternative GM-VFNS approach, 
in the SACOT-$m_T$ scheme, was described in 
Ref.~\cite{Helenius:2018uul}, but cross sections of $J/\psi$ 
or $\psi(2S)$ production have not yet been calculated in that 
scheme.

%%%%%%%%%%%%%%%%%%%%%%%%%%%%%%%%%%%%%%%%%%%%%%%%%%%%%%%%%%%%%%%%%
\section{Results}

\begin{figure*}[t!]
\begin{center}
\includegraphics[width=8.0cm]{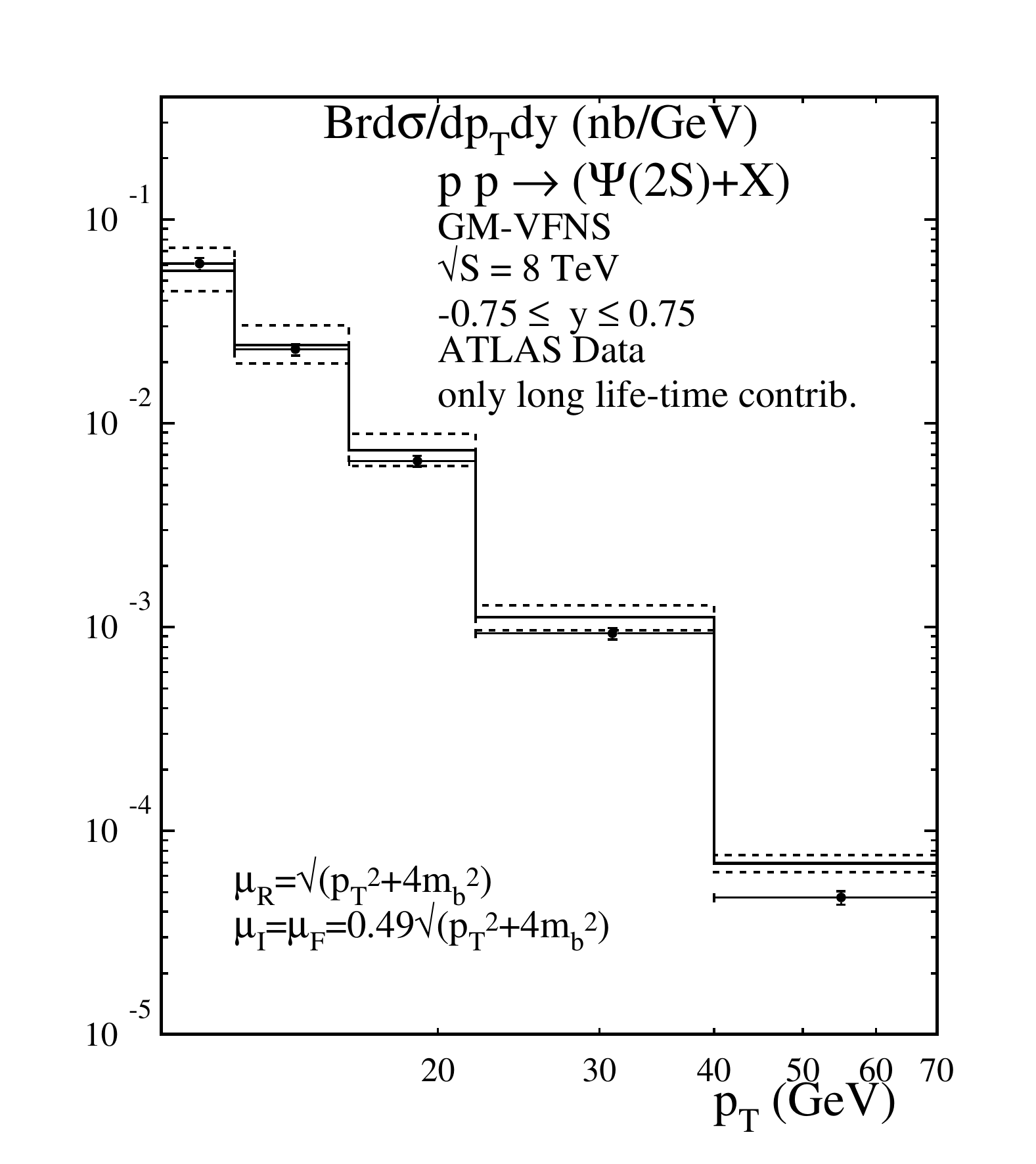}
\includegraphics[width=8.0cm]{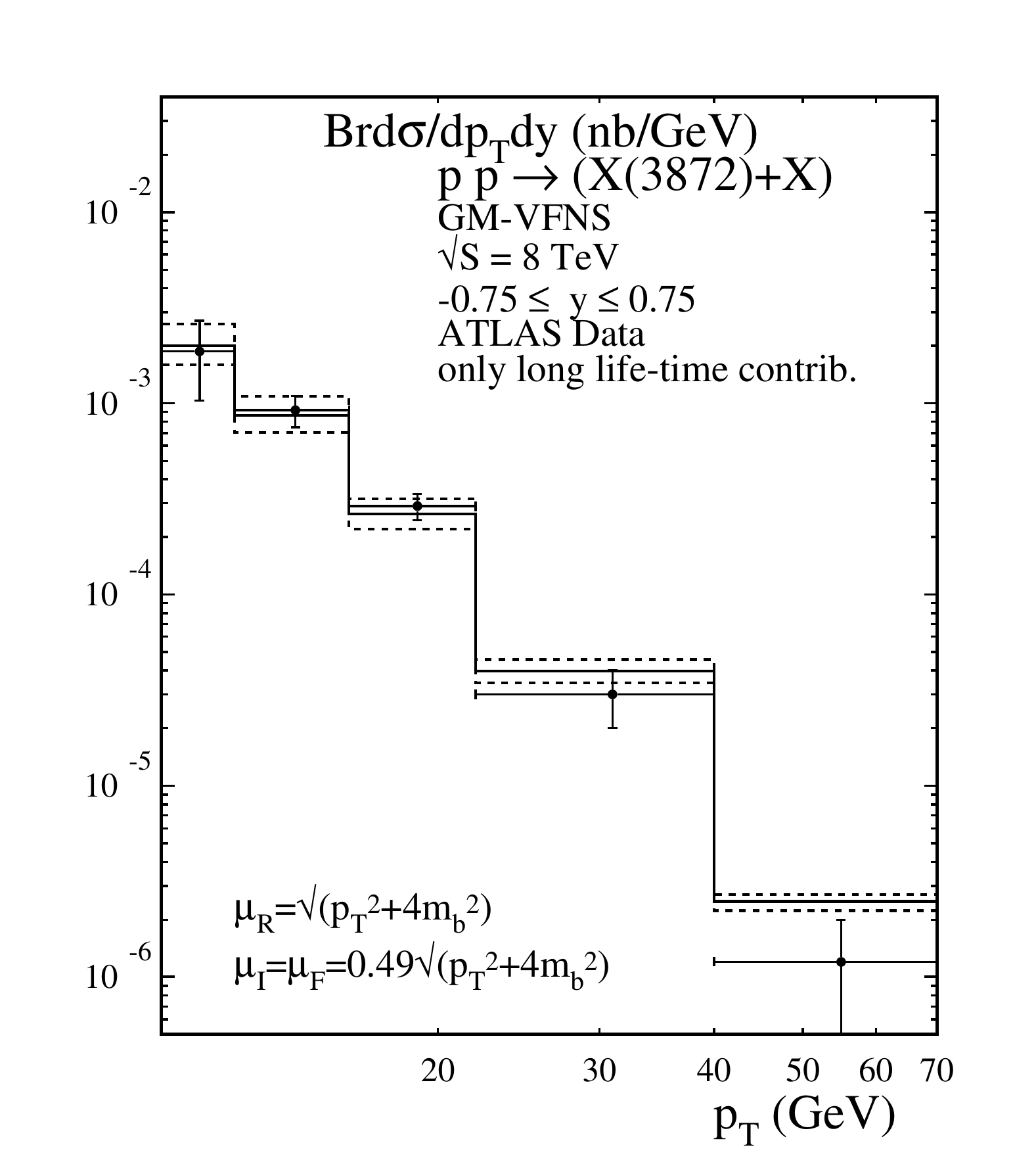}
\end{center}
\caption{
  Double differential cross sections times branching fractions 
  for (a) $\psi(2S)$ (left) and (b) $X(3872)$ (right) nonprompt 
  hadroproduction at $\sqrt{S} = 8$~TeV, averaged over $y$ in the 
  region $|y|\leq 0.75$, as a function of $p_T$. The solid points 
  with error bars represent the ATLAS data \cite{Aaboud:2016vzw}.
  The solid and the upper and lower dashed histograms represent 
  our NLO GM-VFNS predictions for $\xi=1,0.5,2$, respectively.
}
\label{fig:1} 
\end{figure*}

We start by considering $\psi(2S)$ nonprompt hadroproduction.
Its theoretical treatment in the GM-VFNS was described in
Ref.~\cite{Bolzoni:2013tca}. Without any additional assumptions, 
we thus obtain the results for the differential cross section 
$d\sigma/(dp_Tdy)$ at $\sqrt{S} = 8$~TeV, averaged over the rapidity 
range $|y| < 0.75$, which are presented for $10 < p_T < 70$~GeV 
in Fig.~\ref{fig:1}(a) and compared with the ATLAS data 
\cite{Aaboud:2016vzw}. We adopt the choices of renormalization 
and (unified) factorization scales, $\mu_R$ and $\mu_F$, from our 
recent analysis of $B^\pm$-meson hadroproduction at the LHC 
\cite{Benzke:2019usl} by setting $\mu_R = \xi\mu_T$ and $\mu_F = 
0.49\mu_T$, with $\mu_T=\sqrt{p_T^2+4m_b^2}$ and $m_b = 4.5$~GeV,
and varying $\xi$ between 0.5 and 2 about its default value 1 to 
estimate the theoretical uncertainty. We use set CT14nlo  
\cite{Dulat:2015mca} of parton distribution functions (PDFs)
and the $B$-meson fragmentation functions (FFs) from 
Ref.~\cite{Kniehl:2008zza}. As in the ATLAS analysis 
\cite{Aaboud:2016vzw}, we include the $B^+$, $B^0_s$, and 
$\Lambda_b^0$ hadrons and their antiparticles
\cite{Kartvelishvili:PC} taking into account the respective 
$b \to B$-hadron branching fractions as given in 
Ref.~\cite{Zyla:2020zbs}. Here, we select the ATLAS dataset
corresponding to the long-lifetime contribution determined 
by the two-lifetime fit, where the short-lifetime contribution 
is attributed to the $B_c \to \psi(2S) + X$ decay. The data 
corresponding to the one-lifetime fit are quite similar, except 
that the cross sections in the first two $p_T$ bins are slightly 
larger. We observe from Fig.~\ref{fig:1}(a) that the 
agreement between experiment and theory is quite good, except 
for the uppermost $p_T$ bin. 

We now turn to $X(3872)$ nonprompt hadroproduction. We convert 
our results for $\psi(2S)$ nonprompt hadroproduction shown in
Fig.~\ref{fig:1}(a) to the $X(3872)$ case by including 
the $R_{B}^{\text{2L}}$ ratio listed in Eq.~(\ref{eq:rb2l}).
In Fig.~\ref{fig:1}(b), we compare the outcome with the 
respective ATLAS data, again for the two-lifetime fit 
\cite{Aaboud:2016vzw}. The ATLAS data for the single-lifetime fit 
are somewhat larger in the first two $p_T$ bins. We emphasize that 
the value of $R_{B}^{\text{2L}}$ in Eq.~(\ref{eq:rb2l}) comes from the 
same experimental analysis as the cross sections $d\sigma/dp_T$ 
of inclusive $\psi(2S)$ and $X(3872)$ production, albeit from a 
different observable, namely from the ratio of long-lived $X(3872)$  
to long-lived $\psi(2S)$ production rates with additional information 
from the $p_T$ dependencies [see Fig.~4(b) in Ref.~\cite{Aaboud:2016vzw}]. 
Of course, one would like to have this information also from an 
independent experiment---for example from $\psi(2S)$ and $X(3872)$ 
production in $e^+e^-$ annihilation. Unfortunately such information 
is not available yet. 

%%%%%%%%%%%%%%%%%%%%%%%%%%%%%%%%%
%%%%%%%%%%%%%%%%%%%%%%%%%%%%%%%%%%

The quality of agreement between data and theoretical predictions 
is similar for $\psi(2S)$ and $X(3872)$ production. It would be 
interesting to perform a comparative analysis of $\psi(2S)$ and 
$X(3872)$ production in $e^+e^-$ annihilation. After all, the 
$R_B$ ratio should not depend on the production mechanism.

Our results for $X(3872)$ nonprompt hadroproduction 
shown in Fig.~\ref{fig:1}(b) were obtained using the value 
of $R_{B}^{\text{2L}}$ given in Eq.~(\ref{eq:rb2l}). To obtain a 
cross-check, we determine $R_{B}$ by fitting our theory 
prediction to the ATLAS data \cite{Aaboud:2016vzw} for the 
five $p_T$ bins, assuming Gaussian errors. We thus obtain 
\begin{equation}
R_B^{\text{ATLAS}} = \left(3.41 \pm 0.37 \, 
%{+ 0.63\atop- 0.56} \right) \times 10^{-2}
\genfrac{}{}{0pt}{}{+ 0.63}{- 0.56} \right) \times 10^{-2}
\, , 
\label{eq:our-rbsl-atlas}
\end{equation}
where the first error is propagated from the experimental data 
and the second errors are due to the uncertainty of the theory 
prediction from $\xi$ variations as explained above.
This value agrees within errors with the value given in 
Eq.~(\ref{eq:rb2l}) obtained from the two-lifetime fit in 
Ref.~\cite{Aaboud:2016vzw}. Its central value is slightly 
smaller and its error somewhat larger than in Eq.~(\ref{eq:rb2l}). 
We conclude that the determinations of $R_B$ from the two-lifetime 
fit and from the cross-section analyses by ATLAS are consistent 
with each other.

%%%%%%%%%%%%%%%%%%%%%%%%%%%%%%%%%
\begin{figure*}[t!]
\begin{center}
\includegraphics[width=8.0cm]{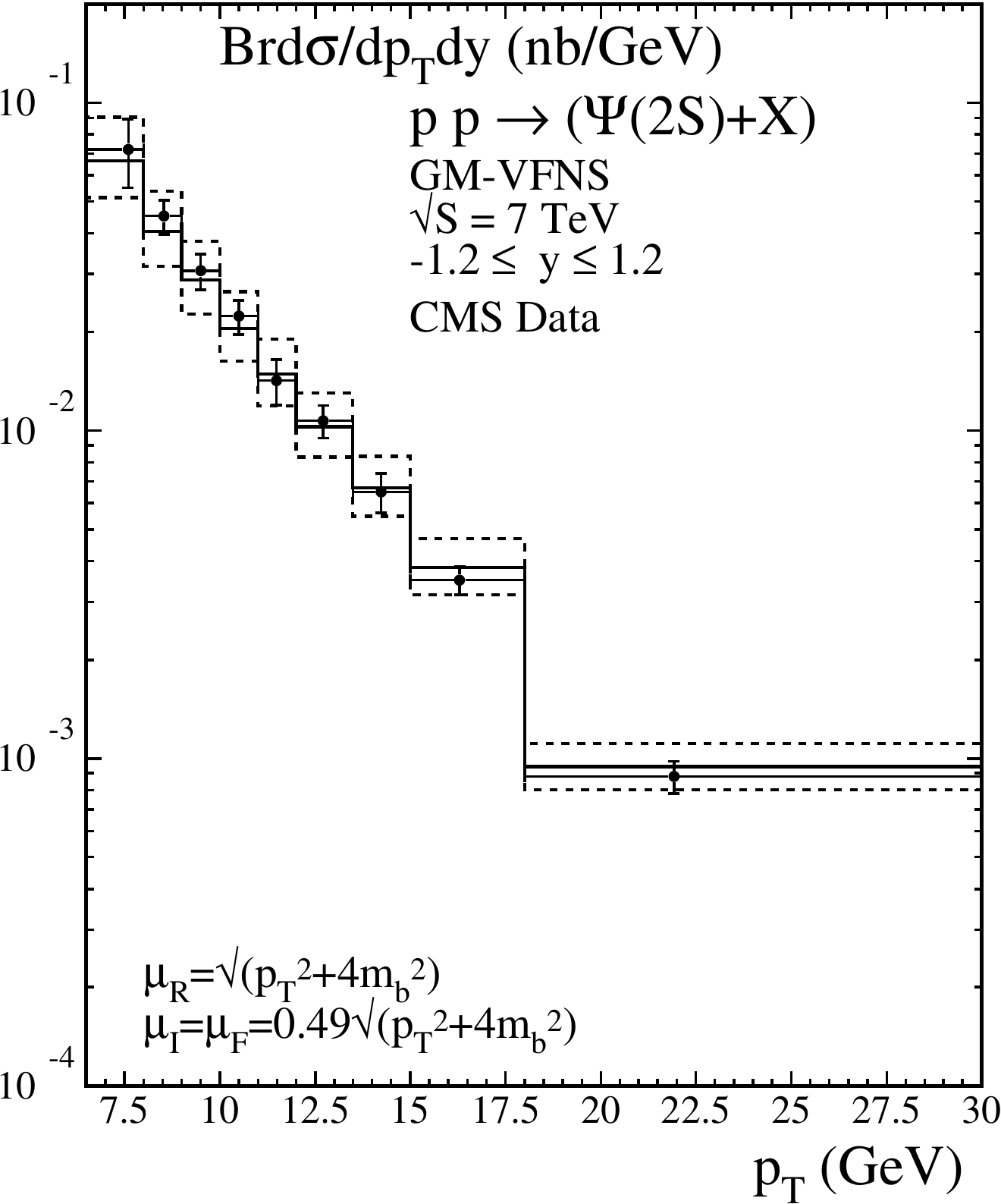}
\end{center}
\caption{%
  Double differential cross section times branching fraction 
  for $\psi(2S)$ nonprompt hadroproduction at $\sqrt{S} = 7$~TeV, averaged
  over $y$ in the region $|y| \leq 1.2$, as a function of $p_T$. The solid
  points with error bars represent the CMS data \cite{Chatrchyan:2011kc}.
  The solid and the upper and lower dashed histograms represent 
  our NLO GM-VFNS predictions for $\xi=1,0.5,2$, respectively.
\label{fig:2} 
}
\end{figure*}
%%%%%%%%%%%%%%%%%%%%%%%%%%%%%%%%%%

To obtain yet another check of our assumption that the $R_B$ 
value in Eq.~(\ref{eq:rb2l}) \cite{Aaboud:2016vzw} is realistic, 
we use it for a comparison with the inclusive cross section 
$d\sigma/dp_T$, integrated over $|y|<1.2$, of nonprompt $X(3872)$ 
production measured by the CMS Collaboration at $\sqrt{S} = 7$~TeV 
\cite{Chatrchyan:2013cld}. In Ref.~\cite{Chatrchyan:2013cld}, this 
cross section was not reported directly, but it can be calculated 
in four $p_T$ bins from the prompt $X(3872)$ cross section times 
branching ratio (Table~7 in Ref.~\cite{Chatrchyan:2013cld})
and the $X(3872)$ nonprompt fraction (Table 6 in 
Ref.~\cite{Chatrchyan:2013cld}). Since both measurements have 
rather large uncertainties, the same is true also for the nonprompt 
cross section, but this should be sufficient for the sake of an 
approximate cross-check. In Ref.~\cite{Chatrchyan:2013cld}, CMS 
used as a benchmark their earlier measurement of the inclusive cross 
section of nonprompt $\psi(2S)$ production, also at $\sqrt{S} = 
7$~TeV and for $|y|<1.2$, but with a different $p_T$ binning 
\cite{Chatrchyan:2011kc}. In Ref.~\cite{Bolzoni:2013tca}, we 
compared these data with our prediction, which came as a continuous 
function in $p_T$.
We now repeat this comparison adopting the very $p_T$ binning from
Ref.~\cite{Chatrchyan:2011kc} and present the outcome in Fig.~\ref{fig:2}.
The good agreement which is evident from Fig.~\ref{fig:2} reassures us
that our prediction of nonprompt $\psi(2S)$ production serves as a useful
starting point for generating a prediction of nonprompt $X(3872)$ production.

%%%%%%%%%%%%%%%%%%%%%%%%%%%%%%%%%
\begin{figure*}[b!]
\begin{center}
  \includegraphics[width=8.0cm]{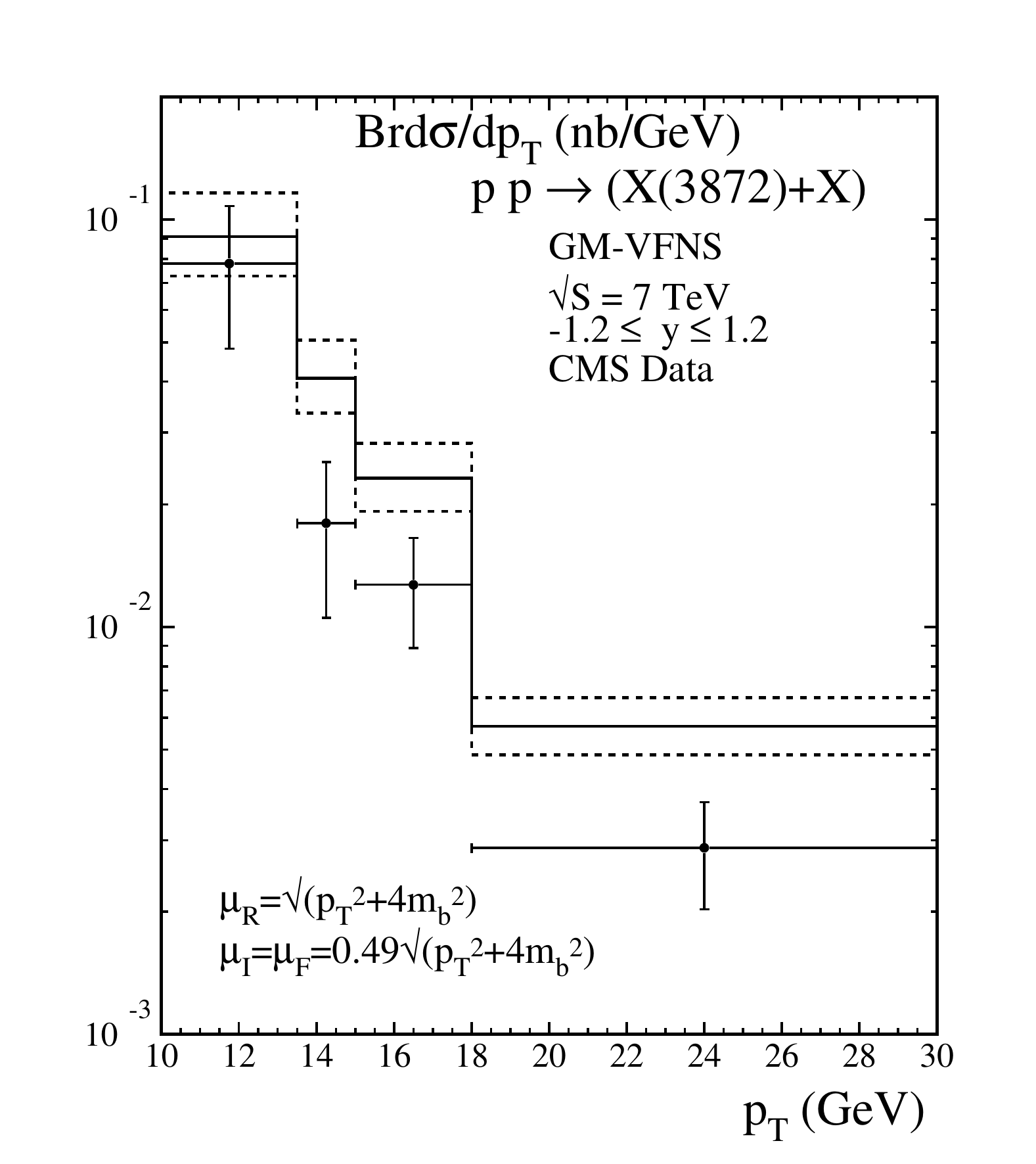}
  \includegraphics[width=8.0cm]{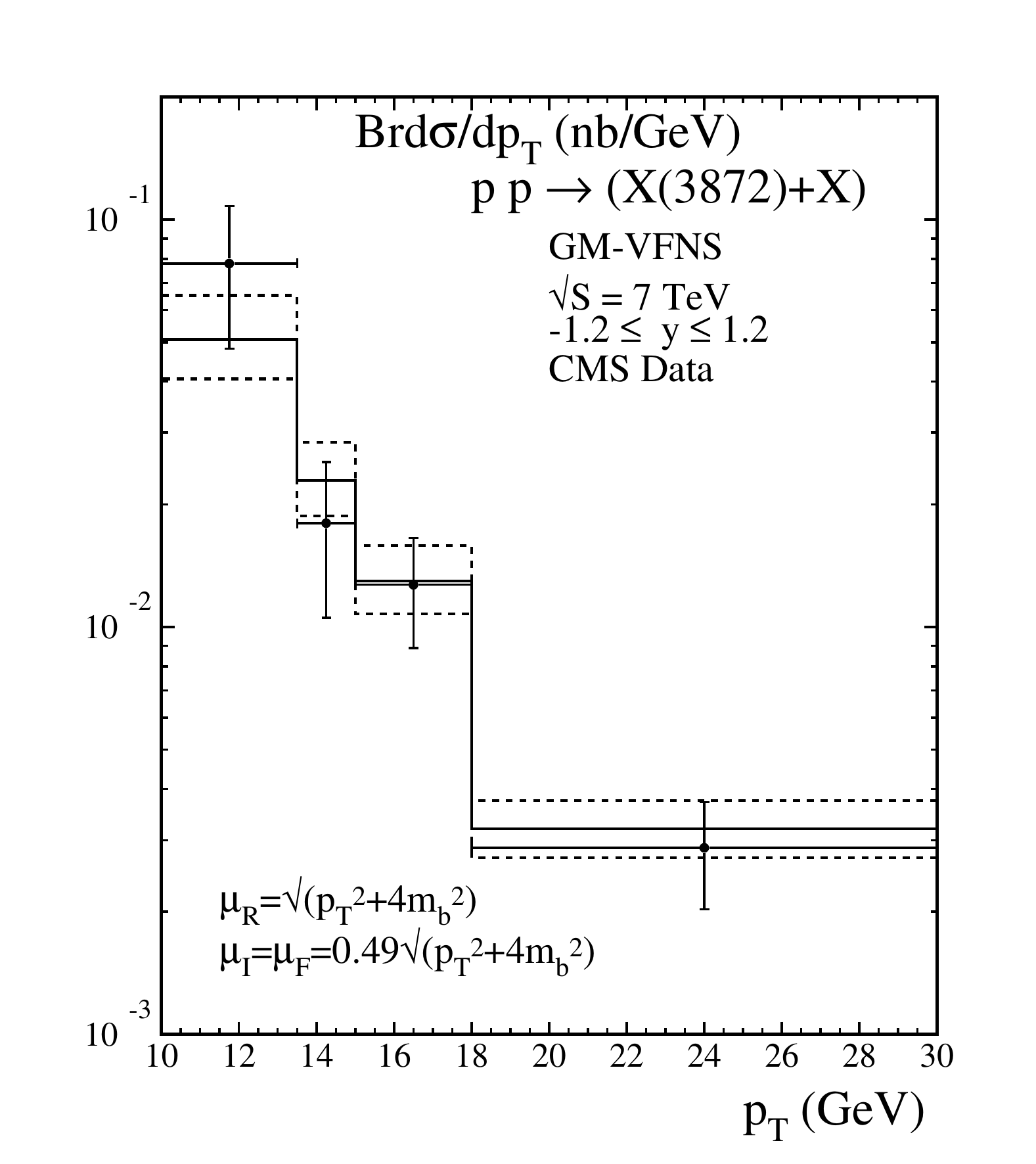}
\end{center}
\caption{
  Differential cross sections times branching fractions for 
  $X(3872)$ nonprompt hadroproduction at $\sqrt{S} = 7$~TeV, 
  integrated over $y$ in the region $|y| \leq 1.2$, as a 
  function of $p_T$. The solid points with error bars are 
  extracted from the CMS data \cite{Chatrchyan:2013cld}. 
  The solid and the upper and lower dashed histograms represent 
  our NLO GM-VFNS predictions for $\xi=1,0.5,2$, respectively. 
  The theoretical predictions were evaluated using the $R_B$ 
  values in (a) Eq.~\eqref{eq:rb2l} (left) and (b) 
  Eq.~\eqref{eq:our-rbsl-cms} (right).
}
\label{fig:3} 
\end{figure*}
%%%%%%%%%%%%%%%%%%%%%%%%%%%%%%%%%%

We thus proceed by adjusting the $p_T$ binning according to
Ref.~\cite{Chatrchyan:2013cld} and including the $R_B$ value from
Eq.~\eqref{eq:rb2l}. In Fig.~\ref{fig:3}(a), the outcome is 
compared with the inclusive cross section of nonprompt $X(3872)$ 
production extracted from Ref.~\cite{Chatrchyan:2013cld} as 
explained above. We find satisfactory agreement within the rather 
large experimental errors, albeit not as good as for the ATLAS 
data \cite{Aaboud:2016vzw} in Fig.~\ref{fig:1}(b). However, our 
predictions do not deviate more than 2 standard deviations from the CMS data
\cite{Chatrchyan:2013cld}. It is obvious that there is no room 
for a discrepancy by a factor between 4 and 8 as claimed in 
Ref.~\cite{Aaboud:2016vzw}. The $\pm10\%$ uncertainty in the 
$R_B$ value in Eq.~\eqref{eq:rb2l}, which is not included in 
Fig.~\ref{fig:3}(a), does not affect this conclusion. 
 
In turn, we may now use the CMS data \cite{Chatrchyan:2013cld} 
for an independent determination of $R_B$, just as we did above 
with the ATLAS data \cite{Aaboud:2016vzw}. We thus find
\begin{equation}
R_B^{\text{CMS}} = \left(1.89 \pm 0.32 \, 
{+ 0.38\atop- 0.33} \right) \times 10^{-2}
\, .
\label{eq:our-rbsl-cms}
\end{equation}
As expected from Fig.~\ref{fig:3}(a), this result is smaller 
than the one in Eq.~(\ref{eq:rb2l}) from the two-lifetime fit 
in the ATLAS publication \cite{Aaboud:2016vzw} and the one in 
Eq.~\eqref{eq:our-rbsl-atlas} from matching our $\psi(2S)$ 
predictions to the ATLAS $X(3872)$ data \cite{Aaboud:2016vzw}.
For a consistency check, we compare the CMS $X(3872)$ data
\cite{Chatrchyan:2013cld} with our predictions evaluated with 
the $R_B$ value fitted to these data in Fig.~\ref{fig:3}(b), 
to find good agreement as expected.

We also simultaneously fit $R_B$ to the nonprompt $X(3872)$ data 
from ATLAS \cite{Aaboud:2016vzw} and CMS \cite{Chatrchyan:2013cld}.
This is possibly problematic because the experimental (first) errors 
of $R_B$ in Eqs.~\eqref{eq:our-rbsl-atlas} and \eqref{eq:our-rbsl-cms} 
from the individual fits do not overlap. In other words, the agreement 
of the underlying ATLAS \cite{Aaboud:2016vzw} and CMS 
\cite{Chatrchyan:2013cld} data, gauged with respect to our universal
NLO GM-VFNS predictions, is marginal. In fact, the combined fit to the 
$N=9$ experimental data points yields a minimum $\chi^2$ of 
$\chi_{\text{min}}^2=14.70$. Following a recommendation by the 
Particle Data Group \cite{Zyla:2020zbs}, we thus rescale the original 
Gaussian error of $0.24$ with the enhancement factor
$\sqrt{\chi_{\text{min}}^2/(N-1)}=1.36$.
We so obtain
\begin{equation}
R_B^{\text{ATLAS+CMS}}  = \left(2.54 \pm 0.33 \, 
{+ 0.49\atop- 0.43} \right) \times 10^{-2}
\, ,
\label{eq:our-rbsl-atlascms}
\end{equation}
where the theoretical (second) error is determined as in
Eqs.~\eqref{eq:our-rbsl-atlas} and \eqref{eq:our-rbsl-cms}.

Finally, it is interesting to compare our results to recent data from the LHCb
Collaboration on the dependence on the charged-particle multiplicity $N$ of
nonprompt $\psi(2S)$ and $X(3872)$ production, with subsequent decays to
$J/\psi\pi^+\pi^-$, in $pp$ collisions at $\sqrt{S}=8$~TeV
\cite{Aaij:2020hpf}.\footnote{%
  In Ref.~\cite{Aaij:2020hpf}, the $X(3872)$ hadron is denoted $\chi_{c1}(2S)$,
  an identification that was found to be disfavored through a dedicated NLO
  NRQCD analysis of prompt $X(3872)$ hadroproduction
  \cite{Butenschoen:2013pxa,Butenschoen:2019npa}.}
The study of the $X(3872)$ to $\psi(2S)$ ratio of nonprompt production cross
sections as a function of $N$ may help to shed light on the nature of the
exotic $X(3872)$ state \cite{Braaten:2020iqw}.
%
%Identifying this ratio with $R_B$, as we do in this paper, we extract from
%Fig.~4 in Ref.~\cite{Aaij:2020hpf} for each interval $[N_{\rm min},N_{\rm max}]$
%with weighted center $N$ the measured value of $R_B$ and compile the results
%in Tab.~\ref{tab:lhcb}.
This ratio, which we identify with $R_B$, as we do in this paper, is presented
in Fig.~4 of Ref.~\cite{Aaij:2020hpf} for five intervals
$[N_{\rm min},N_{\rm max}]$ with weighted centers $N$.
The results \cite{private} are compiled in Table~\ref{tab:lhcb}.
\begin{table}[t!]
\begin{center}
\begin{tabular}{r|r|r|r}
%$N_{\rm min}$ & $N_{\rm max}$ & $N$ & $R_B$ \\[1mm]
$N_{\rm min}$ & $N_{\rm max}$ & $N$ & $R_B\times10^2$ \\[1mm]
\hline
\rule[-1mm]{0mm}{5mm}
%0 & 40 & 30.8 & $0.0265459 \pm 0.0132296 \pm 0.00203169$
%\\
%40 & 60 & 50.3 & $0.0316677 \pm 0.00595951 \pm 0.00238706$
%\\
%60 & 80 & 69.3 & $0.0346934 \pm 0.00706946 \pm 0.00259032$
%\\
%80 & 100& 88.8 & $0.0443463 \pm 0.0119505 \pm 0.00322719$
%\\
%100 & 200 & 118.3 & $0.0533022 \pm 0.0158538 \pm 0.00386698$
0 & 40 & 30.8 & $2.65\pm1.32\pm0.20$
\\
40 & 60 & 50.3 & $3.17\pm0.60\pm0.24$
\\
60 & 80 & 69.3 & $3.47\pm0.71\pm0.26$
\\
80 & 100& 88.8 & $4.43\pm1.20\pm0.32$
\\
100 & 200 & 118.3 & $5.33\pm1.59\pm0.39$
\end{tabular}
\caption{
$R_B$ for five different bins of charged-particle multiplicity $N$ as measured 
  by the LHCb Collaboration in $pp$ collisions at $\sqrt{S}=8$~TeV.
  The first and second errors are the uncorrelated and correlated
  uncertainties, respectively.
  The data are presented in Fig.~4 of Ref.~\cite{Aaij:2020hpf}. 
\label{tab:lhcb}
}
\end{center}
\end{table}
We observe from Table~\ref{tab:lhcb} a slight increase of $R_B$ with $N$,
which is, however, not statistically significant.
As discussed in Ref.~\cite{Aaij:2020hpf}, a linear fit to these data 
points, without considering the correlated systematic uncertainty, 
gives a positive slope that is consistent with zero within 1.6 
standard deviations.
For this reason, it appears to be reasonable to fit a horizontal line to these
data points.
Specifically, we perform three fits using the uncorrelated uncertainties:
one to the central data points, and one each to the central data points
shifted up and down by their correlated uncertainties.
We then take the upward and downward shifts of the central value to be the
error due to the correlated uncertainties, which, in want of the correlation
matrix, represents a conservative estimate.
%In this way, we obtain an average weighted by the uncorrelated
%uncertainty on each data point.
This yields
\begin{equation}
  R_B^{\text{LHCb}} = (3.48 \pm 0.39 \pm 0.26) \times 10^{-2}\, ,
  \label{eq:lhcb}
\end{equation}
where the first and second errors stem from the uncorrelated and correlated
uncertainties, respectively.
The fit to the unshifted data points yields a $\chi^2$ per degree of freedom as
low as $\chi^2/4=0.67$, providing {\it a posteriori} a convincing justification
for the zero-slope hypothesis. 
The value in Eq.~\eqref{eq:lhcb} is in excellent agreement with the
two-lifetime fit from ATLAS \cite{Aaboud:2016vzw}.

In Ref.~\cite{Braaten:2020iqw}, the LHCb data of prompt and nonprompt
$X(3872)$ hadroproduction \cite{Aaij:2020hpf} were jointly fitted to a model,
assuming that $R_B$ is independent of $N$, as we do.
Among other things, this yielded
\begin{equation}
  R_B^{\text{\cite{Braaten:2020iqw}}} = (3.24 \pm 0.29) \times 10^{-2}\, ,
  \label{eq:braaten}
\end{equation}
which is consistent with our fit result in Eq.~\eqref{eq:lhcb}.

For the reader's convenience, we summarize in Table~\ref{tab:Rb} the various
values for $R_B$ discussed in this paper.
\begin{table}[h!]
\begin{center}
\begin{tabular}{l|l|l}
\rule[-2mm]{0mm}{8mm}
Name & $R_B \times 10^2$ & Source 
\\
\hline
\rule[-2mm]{0mm}{8mm}
$R_B^{\text{\cite{Artoisenet:2009wk}}}$ & $18 \pm 8$ & Extracted from CDF~II data
\cite{Bauer:2004bc} in Ref.~\cite{Artoisenet:2009wk} 
\\
\rule[-1mm]{0mm}{5mm}
$R_B^{\text{2L}}$ & $3.57 \pm 0.348$ & ATLAS \cite{Aaboud:2016vzw}
\\
\rule[-1mm]{0mm}{5mm}
$R_B^{\text{LHCb}}$ & $3.48 \pm 0.39 \pm 0.26$ & Extracted from LHCb data
\cite{Aaij:2020hpf} here
\\
\rule[-1mm]{0mm}{5mm}
$R_B^{\text{\cite{Braaten:2020iqw}}}$ & $3.24 \pm 0.29$ & Extracted from LHCb data
\cite{Aaij:2020hpf} in Ref.~\cite{Braaten:2020iqw}
\\
\rule[-1mm]{0mm}{5mm}
$R_B^{\text{ATLAS}}$ & $3.41 \pm 0.37 \, {+ 0.63\atop- 0.56}$ & Our fit to ATLAS data \cite{Aaboud:2016vzw}
\\
\rule[-1mm]{0mm}{5mm}
$R_B^{\text{CMS}}$ & $1.89 \pm 0.32 \, {+ 0.38\atop- 0.33}$ & Our fit to CMS data 
\cite{Chatrchyan:2013cld}
\\
\rule[-1mm]{0mm}{5mm}
$R_B^{\text{ATLAS+CMS}}$ &  $2.54 \pm 0.33 \, {+ 0.49\atop- 0.43}$ & Our joint fit to ATLAS \cite{Aaboud:2016vzw} and CMS \cite{Chatrchyan:2013cld} data
\end{tabular}
\caption{
Summary of the different $R_B$ determinations discussed in this article. 
\label{tab:Rb}
}
\end{center}
\end{table}

%%%%%%%%%%%%%%%%%%%%%%%%%%%%%%%%%%%%%%%%%%%%%%%%%%%%%%%%%%%%%%%%%%%%%

\section{Summary} 

We updated our prediction of the inclusive cross section of 
nonprompt $\psi(2S)$ hadroproduction at NLO in the GM-VFNS 
\cite{Bolzoni:2013tca} and validated it by comparison with ATLAS 
\cite{Aaboud:2016vzw} and CMS \cite{Chatrchyan:2011kc} data.
From this, we obtained an analogous prediction for nonprompt 
$X(3872)$ hadrons by including the appropriate ratio $R_B$ of 
branching fractions. In turn, this enabled us to determine $R_B$ 
by fitting to ATLAS \cite{Aaboud:2016vzw} and CMS 
\cite{Chatrchyan:2013cld} data of nonprompt $X(3872)$ production.
This also provided us with a useful test bed to assess 
determinations of $R_B$ by other authors
\cite{Aaboud:2016vzw,Artoisenet:2009wk,Aaij:2020hpf,Braaten:2020iqw}.
Our findings support the results for $R_B$ obtained by ATLAS
\cite{Aaboud:2016vzw} and LHCb \cite{Aaij:2020hpf,Braaten:2020iqw}, which
undershoot a previous result \cite{Artoisenet:2009wk} by a factor of about
1/5.

\section*{Acknowledgments} 

We thank J.~M. Durham and M.~Winn for useful communications regarding
Ref.~\cite{Aaij:2020hpf} and J.~M. Durham for providing us with the data
points listed in Table~\ref{tab:lhcb} in numerical form.
We are grateful to E.~Braaten for a useful communication regarding
Ref.~\cite{Braaten:2020iqw}.
The  work of B.~A.~K was supported in part by the German Federal Ministry for
Education and Research BMBF through Grant No.\ 05H18GUCC1 and by the German
Research Foundation DFG through Grants No.\ KN~365/13-1 and No.\ KN~365/14-1
within Research Unit FOR~2926 {\it Next Generation Perturbative QCD for Hadron
Structure: Preparing for the Electron-Ion Collider}.
The work of I.~S. was supported in part by the French National Centre for
Scientific Research CNRS through IN2P3 Project GLUE@NLO.

%%%%%%%%%%%%%%%%%%%%%%%%%%%%%%%%%%%%%%%%%%%%%%%%%%%%%%%%%%%%%%%%%%%%%
%%%%%%%%%%%%%%%%%%%%%%%%%%%%%%%%%%%%%%%%%%%%%%%%%%%%%%%%%%%%%%%%%%%%%

%%%%%%%%%%%%%%%%%%%%%%%%%%%%%%%%%%%%%%%%%%%%%%%%%%%%%%%%%%%%%%%%%%%%%
%%%%%%%%%%%%%%%%%%%%%%%%%%%%%%%%%%%%%%%%%%%%%%%%%%%%%%%%%%%%%%%%%%%%%

\end{document}